\documentclass{PoS}
\usepackage{subcaption}

\title{On a four-loop form factor in N=4}

\ShortTitle{On a four-loop form factor in N=4}

\author{\speaker{Rutger Boels},  Bernd A. Kniehl \thanks{supported by the German Science Foundation (DFG) within the Collaborative Research Center 676 ``Particles, Strings and the Early Universe"}\\
        II. Institut f\"ur Theoretische Physik, Universit\"at Hamburg\\ Luruper Chaussee 149, D- 22761 Hamburg, Germany\\
        E-mail: \email{Rutger.Boels@desy.de} and \email{Bernd.Kniehl@desy.de}}

\author{Gang Yang \\% \thanks{supported by a DFG grant in the framework of the SFB 647 ``Space-Time-Matter"}\\
	CAS Key Laboratory of Theoretical Physics, Institute of Theoretical Physics, \\Chinese Academy of Sciences, Beijing 100190, China\\
	E-mail: \email{yangg@itp.ac.cn}}

\abstract{We report on progress toward computing a four-loop supersymmetric form factor in maximally supersymmetric Yang-Mills theory. A representative example particle content from the involved supermultiplets is a stress-tensor operator with two on-shell gluons. In previous work, the integrand for this form factor was obtained using color-kinematic duality in a particularly simple form. Here the result of applying integration-by-parts identities is discussed and cross-checks of the result is performed. Rational IBP relations and their reduction are introduced as a potentially useful aide.}

\FullConference{Loops and Legs in Quantum Field Theory\\
		24-29 April 2016\\
		Leipzig, Germany}

\begin{document}

\section{Introduction}

Difficulty in perturbative quantum field theory can be measured along the axes of numbers of loops and numbers of legs in a typical problem to be computed. Entangled with these is the issue of the number of scales in a typical problem: the more scales (such as more legs or loops), the harder the problem. For instance, single loop problems with many legs as well as multiple scales can nowadays be solved routinely, modulo numerical instabilities and issues. For practical collider applications though, the frontier of development is for on-shell scattering amplitudes found already at two loops, with five or more on-shell massless particles. This holds even before taking into account the important issue of obtaining a finite cross-section after cancelling of UV and IR divergences. This motivates the push for more loops and legs. 

An important tool for pushing development in perturbative quantum field theory is the use of toy models. A particularly important example of such a model is the unique maximally supersymmetric Yang-Mills theory in four dimensions. This has played an important role as a testing ground for new ideas and methods in the recent past \cite{Witten:2003nn} and is interesting in its own right for the relation of its planar sector to a string theory description in the AdS/CFT correspondence \cite{Maldacena:1997re}. In the context of the latter some quantities have a (partly conjectural) exact description. For the present context, an important example of this is the planar light-like cusp anomalous dimension, for which an exact integral equation exists \cite{Beisert:2006ez}. The planar limit provides particularly nice simplifications. In this limit the number of colors, $N_c$, in a $SU(N_c)$ gauge theory is taken to infinity. In the physical theory of the strong interactions however, this parameter is three and non-planar effects are physically relevant. Also from the formal point of view the value of the first non-planar corrections to the cusp anomalous dimension is highly interesting, related to string loop corrections in the AdS space. 

A second, but no less important reason to be interested in the light-like cusp anomalous dimension is the fact that it is a universal quantity that is ubiquitous in QFT computations. It appears for instance in the universal exponentiation of IR divergences in dimensional regularisation and gets its name from the computation of the light-like cusped Wilson line \cite{Korchemsky:1985xj, Korchemsky:1987wg}. This function is usually mixed in with additional structure in a given scattering amplitude or form factor. The exception to this is the simplest, Sudakov form factor. This consists of a member of the stress-tensor multiplet (i.e. a local gauge invariant operator) and two members of the on-shell gluon multiplet that is unique in $\mathcal{N}=4$ SYM. This is a single-scale problem, and it can be shown that maximal supersymmetry factors off a universal tree-level form factor that contains all dependence on external polarizations as well as the dependence on the gluon color quantum numbers. Hence to any loop order
\begin{equation}
F^{(l)} = F^{(0)}  (-q^2 )^{-l \epsilon } {\tilde F}^{(l)} 
\end{equation}
holds, where $q^2= (p_1 + p_2)^2$ and $D=4-2 \epsilon$, and ${\tilde F}^{(l)}$ is now a dimensionless function of $g_{\textrm{ym}}$, $N_c$ and $\epsilon$. It is related to the cusp anomalous dimension by exponentiation \cite{Mueller:1979ih, Collins:1980ih, Sen:1981sd, Magnea:1990zb},
\begin{equation}\label{eq:expIR}
{\tilde F}^{(l)} =  \exp\bigg( {\sum_l \frac{-g^{2l}\gamma_{\textrm{cusp}}^{(l)} }{(2l \epsilon)^2}  } + {\cal O}(\epsilon^{-1}) \bigg) \, .
\end{equation}
Therefore at fixed order in perturbation theory one can compute the correction to the light-like cusp anomalous dimension to that order in perturbation theory. The possible algebraic color invariants that can appear are determined from all possible contractions of structure constants (see e.g. \cite{Boels:2012ew} for results up to eight loops). Analysing their $N_c$ dependence shows that the first non-planar correction appears at four loops. It is therefore a particular interesting and challenging problem to consider four-loop form factors, which is the focus of this paper.

The computation of the Sudakov form factor has a long history both within $\mathcal{N}=4$ as well as in more physical theories \cite{Mueller:1979ih, Collins:1980ih, Sen:1981sd}. The first two-loop correction in $\mathcal{N}=4$ appeared already in  \cite{vanNeerven:1985ja}. The three-loop correction in QCD was studied in a series of papers starting with a basis of integral after IBP reduction  \cite{Gehrmann:2006wg}, through numerical analysis \cite{Baikov:2009bg} to analytic integration  \cite{Lee:2010cga}, with an important cross-check in \cite{Gehrmann:2010ue}. In \cite{Henn:2013wfa} the form factor in $\mathcal{N}=4$ to the three-loop order was studied. See also \cite{Henn:2016men} for recent exciting progress at four loops in QCD.  In \cite{Boels:2012ew} the first simple form of the integrand of the four-loop form factor in $\mathcal{N}=4$ was published using color-kinematic duality \cite{Bern:2008qj, Bern:2010ue, Bern:2012uf}. This was followed by the IBP reduction of its integrals in  \cite{Boels:2015yna}, to which the reader is also referred to for details of part of the work presented here. Interesting ideas for evaluating the integrals based on dimensional recurrences  \cite{Tarasov:1996br} and advanced integral analysis have appeared in \cite{vonManteuffel:2014qoa}  and \cite{vonManteuffel:2015gxa}.

\section{Integration-by-parts identities and their solution}
A crucial step in  state-of-the-art computations in quantum field theory is the use of integration-by-parts identities \cite{Chetyrkin:1981qh, Tkachov:1981wb} to simplify a given problem. Any integrand-generating technique such as Feynman graphs eventually produces a sum over integrals of the type
\begin{equation}\label{eq:deffeynmanint}
I(a_1, \ldots, a_n) \equiv \int d^D l_1 \ldots d^Dl_L  \left(1/D_1\right)^{a_1} \ldots  \left(1/D_n\right)^{a_n} ,
\end{equation}
where $D_i$'s are typically inverse propagators.
These integrals are not all indepedent, but obey linear identities that follow from 
\begin{equation}
0 = \int d^D l_1 \ldots d^Dl_L  \,\, \frac{\partial}{\partial l_i^{\mu}} \, X \,,
\end{equation}
for any derivative with respect to a loop momentum. This type of identity leads to a large set of linear equations, which can be solved by Gaussian elimination, at least in principle. Since the matrix involved is non-invertible, the solution takes the form of an expression of all integrals in a given set in terms of a small subset. This smaller subset can be chosen to consist of the simplest integrals in the class. The latter implies there is a choice of basis involved, which is to be supplied by the user. This is the essence of the Laporta algorithm \cite{Laporta:2001dd}. 

Many private and public implementations of this approach exist, such as {\tt AIR} \cite{Anastasiou:2004vj}, {\tt FIRE} \cite{Smirnov:2008iw, Smirnov:2013dia, Smirnov:2014hma} and {\tt Reduze} \cite{2010CoPhC.181.1293S, vonManteuffel:2012np}.  See {\tt LiteRed} \cite{Lee:2012cn, Lee:2013mka} for an alternative approach to IBP reduction. For the four-loop form factor under study here only {\tt Reduze} \cite{vonManteuffel:2012np} gave an answer in large but finite time. The result is given in an explicit set of reduction form of integrals in the form factor. Some statistics on master integral numbers as a function of numerator power $s$ is given in Table \ref{table:countingReduze}, see \cite{Boels:2015yna} for further details.

\begin{table}[ht]

\caption{Master integral statistics of obtained IBP reduction.}
\begin{subtable}{.5 \linewidth}
\caption{planar form factor} 
\centering
\begin{tabular}{c | c c c} % centered columns (4 columns)
\hline %inserts double horizontal lines
\# props & $s=0$ & s=1 &$ s=2$ \\ [0.5ex] % inserts table
%heading
\hline % inserts single horizontal line
12 & 8 & 6 & 0 \\ % inserting body of the table
11 & 18 & 2 & 0  \\
10 & 43 & 9 & 0 \\
9  & 49 & 1 & 0 \\
8  & 51 & 4  & 1  \\
7 & 25 & 0 & 0 \\
6 & 8 & 0 & 0 \\
5 & 0 & 0 & 0  \\ 
\hline %inserts single line
sum & 203 & 22 & 1  \\ [1ex] % [1ex] adds vertical space
\end{tabular}
\end{subtable}%
\begin{subtable}{.5\linewidth}
\caption{non-planar form factor} 
\centering
\begin{tabular}{c | c c c} % centered columns (4 columns)
\hline %inserts double horizontal lines
\# props & $s=0$ & s=1 &$ s=2$ \\ [0.5ex] % inserts table
%heading
\hline % inserts single horizontal line
12 & 10 & 10 & 1 \\ % inserting body of the table
11 & 13 & 3 & 0  \\
10 & 34 & 10 & 0 \\
9  & 29 & 1 & 0 \\
8  & 32 & 3  & 1  \\
7 & 13 & 0 & 0\\
6 & 7 & 0 & 0 \\
5 & 1 & 0 & 0  \\ 
\hline %inserts single line
sum & 139 & 27 & 2  \\ [1ex] % [1ex] adds vertical space
\end{tabular}

\end{subtable}
\label{table:countingReduze}
\end{table}

\section{Comparison to counting masters using algebraic methods}

The number of master integrals may be counted by exploring only the analytic structure of the integral topology, without explicitly solving the IBP relations.  This method was proposed in  \cite{Lee:2013hzt}, based on earlier work \cite{Baikov:2005nv}.  The basic idea is that, for a given topology of $m$ propagators, the number of master integrals can be obtained by counting the number of proper critical points of the sum of first and second Symanzik polynomials
\begin{equation}
G(\vec\alpha) = U(\vec\alpha) + F(\vec\alpha) \,,
\end{equation} 
where the proper critical points are defined by
\begin{equation}
\label{eq:critical-point-condition}
\frac{\partial G}{\partial \alpha_i} = 0 \ \ \ (i = 1, \ldots, m) \qquad
\mbox{and} \qquad G \neq 0.
\end{equation}
Such critical points can be counted efficiently using the Gr\"obner basis technique. This procedure has been implemented in the {\tt Mint} package  \cite{Lee:2013hzt}.

We have applied this algorithm to the four-loop form factor integrals,  swapping in different approaches to perform some of the steps that the {\tt Mint} package cannot perform in its current incarnation.  We refer the reader to \cite{Boels:2015yna, Lee:2013hzt} for further details. The obtained masters are classified in Table \ref{table:counting_Mint} according to the number of propagators and the power of propagators. A crosscheck with the reduction of {\tt Reduze} shows that all 280 master integrals are independent. 
\begin{table}[t]
\caption{Master integral statistics of 280 Mint basis.}
\centering
\begin{tabular}{c | c c c c c c c c} % centered columns (4 columns)
\hline %inserts double horizontal lines
\# props & 5 & 6 & 7 & 8 & 9 & 10 & 11 & 12 \\
\hline
\textrm{all simple } & 1 & 8 & 25 & 48 & 52 & 58 & 32 & 20 \\
\textrm{simple + one double} & 0 & 0 &  1 &   5 &  1  &  12 &  3  & 14 \\
\hline %inserts single line
%sum &  &  &  \\ [1ex] % [1ex] adds vertical space
\end{tabular}
\label{table:counting_Mint}
\end{table} 
Interestingly, while the counting based on the {\tt Mint} method indeed provides a set of independent basis integrals, we find that the reduction of {\tt Reduze} tends to include more basis integrals. This discrepancy may be due to extra relations beyond IBP, or possible subtle critical points at infinity which are not yet resolved by {\tt Mint} method. It would be very interesting to understand this point. In any case, we expect that the {\tt Mint} basis should provide a lower bound of the number of independent basis. Since the method based on {\tt Mint} only relies on the topologies of the given integrals and applies to arbitrary numerators, the results are expected to apply to any theory, including QCD.

\section{Rational IBP relations}
A particular problem that arises from the explicit reduction is the  expansion in the dimensional regularisation parameter $\epsilon$. Typically a topology within the form factor is given as a sum over master integrals as
\begin{equation}
\textrm{top}_i = \sum c_{i,j} \textrm{mast}_j \,,
\end{equation}
where the coefficients $c_{i,j} $ are functions of $\epsilon$. Since the four-loop topologies are expected in general to diverge as $\frac{1}{\epsilon^8}$ (see equation \ref{eq:expIR}) and for the four-loop cusp the term of order $\frac{1}{\epsilon^2}$ is needed, both prefactors as well as master integrals need to be calculated to high order in the $\epsilon$ expansion. Although the prefactor has an exact expansion, due to mixing of the two expansion the precision in each of the expansion coefficients in the master integral is important. We note that in many cases the coefficients $c_{i,j}$ are such that the integrals $ \textrm{mast}_j$ have to be expanded to the seventh order, regardless of at which order the expansion of these master integrals starts. As a rule of thumb, the coefficients of the prefactor expansion increase an order of magnitude in size with each additional expansion step, making the mixing highly non-trivial. One potential resolution to this problem is to study IBP relations which only contain rational numbers and not $\epsilon$.

One way to obtain such a set will be explained here. All relations with rational coefficients will be obtained for a set of form factor integrals for which a general IBP reduction with dimension dependent conditions has been found previously. In essence, we find the rational relations contained in a given IBP reduction. There are other ways of aiming for such a set ({\tt Reduze} offers a choice of dimension free IBP relations for instance), but using an already available IBP reduction will almost certainly be faster. Even more generally, one could employ the strategy below essentially unchanged to use integer numerics in IBP reductions. This is related to but simpler than the algorithm proposed in \cite{vonManteuffel:2014ixa}. Note that the strategy there aims to yield the full reduction, not just a rational reduction.

\subsection{Algorithm}
The parameter $q^2$ can be set to a fixed value without loss of generality. The only variable for the integrals left is then the dimensional regularisation parameter $\epsilon$. Proceed as follows:

\begin{itemize}
\item write the IBP reduction for all of the integrals in the given set.
\item first remove pairs of integrals: those that have the same IBP reduction. This step requires a minor basis choice (see below). 
\item one is left with $a$ integrals, expressed in terms of $b$ master integrals. Turn this into a ($b$ by $a$) matrix. Its coefficients dependent only on $\epsilon$. 
\item evaluate this matrix for random integer values of $\epsilon$ $x$ times, such that $x b >> a$. 
\item construct the natural $(x \cdot b$ by $a$) matrix by adjoining rows. This matrix is purely rational by construction.
\item compute the null-space of this matrix by computer algebra: these are the sought-for relations.
\item check these relations by using the full dimension-dependent IBP reduction.
\end{itemize}

To obtain an IBP reduction, one always needs to choose an ordering. We choose to order (partially) first by numerator power, then by propagator power and finally by topology number. Lower values are considered simpler. Rearranging the obtained relations according to the ordering (most difficult first) and row reducing the resulting matrix allows to read off an explicit rational IBP reduction with the chosen ordering. 
 
Note that the step removing the duplicates is strictly speaking not essential. Typically, these duplicate cases involve graph symmetries. In cases with many graph symmetries this step is necessary to reduce matrices to a manageable size in the remaining parts of the algorithm. For instance, for all $10$ line integrals across the $34$ topologies, there are $11 \cdot 12 \cdot 34= 4488$ different integrals. After duplicate removal, there are only $389$. These obey $31$ additional rational relations found by the above algorithm, giving a set of $358$ (rational IBP) master integrals with $10$ lines. 

\subsection{Application of rational IBPs to color-planar part of the form factor at four loops}
We have generated all rational relations between the integrals for all of the $34$ topologies. An explicit map to master integrals was obtained using the choice outlined above. This was applied first to the color-planar part of the form factor proportional to $N_c^4$. An immediate problem now is that the result contains, in general, a very large number (typically $>500$) of rational master integrals with, again in general, very complicated looking fractional coefficients (denominators and numerators with many digits). This is likely due to a non-optimal basis choice. 

Some ad-hoc rules to find a better basis choice are:
\begin{itemize}
\item reduction rules with a very high (>125) number of different master integrals must be discarded.
\item integrals from topologies with vanishing color-factor should not appear in the end result. Use duplication relations to eliminate these.
\item any replacement which eliminates several integrals at once out of the sum over all topologies is good.
\item integrals with the largest rational fractions should be reverted first.  
\end{itemize}
One can scan algorithmically for good replacements by selecting a number of integrals and checking in which reduction they appear. Then, one can check the coefficients to test if reverting this relation will decrease the number of integrals in the full result. Taking into account the relations with high numbers of rational master integrals can in rare cases significantly reduce resulting form factor sizes: this happens for instance when two of these relations can be combined into a simpler one\footnote{In the one case known, the two high number of integral expansions contained a number of terms which differed by one unit.}. 

Applying these rules results in a new form of the color-planar part of the four-loop form factor in $\mathcal{N}=4$ with order $\sim 70$ integrals, with coefficients which are small ($1<|x|<100$) integers. The size of the rational IBP reduction is such that it can be used on a laptop, instead of having to be kept and manipulated on a large cluster. This is conducive to experimentation. A drawback is that the remaining set of master integrals now contains quite a number of integrals with quadratic numerators, which are known to be hard to integrate. Hence this approach may not be very fruitful for direct integration without further input. This result is certainly more general: when using or aiming for rational IBP relations, there is no a-priori way of determining the 'power' of the reduction: there is no guarantee the computational complexity of a problem will decrease by solving the set of rational IBP relations.

\section{Conclusion}
In this contribution progress has been discussed toward computing the four-loop cusp anomalous dimension in the special toy model theory of $\mathcal{N}=4$ super Yang-Mills. Current status is that a basis of master integrals after IBP reduction has been obtained, which was cross-checked using algebraic methods, and a counting from the latter methods was argued to apply to the QCD case as well. Significant challenges remain, especially toward integration. A secondary approach through the construction and solution of rational IBP relations was described. For this the known IBP reduction was used to obtain a sub-reduction. Although these techniques yield results which are much more easily manipulated, the reduction in computational complexity is certainly less than with the use of full IBP relations and will require more input for integration. Still, they represent an interesting direction to pursue further, especially for the more complicated classes of integrals which appear in the QCD form factor.

\acknowledgments
The authors would like to thank the organisers of the Loops and Legs 2016 workshop for the opportunity to present this work. It is a pleasure to thank Andreas von Manteufel, Volodya Smirnov, Johannes Henn and Sven-Olaf Moch for discussions. This work was supported by the German Science Foundation (DFG) within the Collaborative Research Center 676 "Particles, Strings and the Early Universe".

\bibliographystyle{JHEP}

\bibliography{4LoopNPCuspNotes}

\end{document}